\documentclass[journal=jacsat,manuscript=article]{achemso}

\usepackage[version=3]{mhchem} 
\usepackage{color}  

\author{Emiliano Poli}
\email{epoli@ictp.it}
\author{Kwang H. Jong}
\author{Ali Hassanali}
\email{ahassanali@ictp.it}
\affiliation[CMPS]
{Condensed Matter Statistical Physics Department, The Abdus Salam International Center for Theoretical Physics, Strada Costiera, Trieste}

\title{Charge Transfer as a Ubiquitous Mechanism in Determining the Negative Charge at Hydrophobic Interfaces}

\abbreviations{IR,NMR,UV}

\begin{document}



\begin{abstract}
The origin of the apparent negative charge at hydrophobic-water interfaces has fueled one of the biggest debates in physical chemistry for several decades. The most common interpretation given to explain this observation is that negatively charged hydroxide ions (OH-) bind strongly to the interfaces. Here, using first principles calculations of the air-water and oil-water interfaces consisting of thousands of atoms, we unravel a mechanism that does not require the presence of OH-. We show that small amounts of charge transfer along hydrogen bonds and asymmetries in the hydrogen bond network, associated with local topological defects can lead to the accumulation of negative surface charge at both interfaces. For water near oil, we show that there is also some spillage of electron density into the oil leaving it negatively charged. The surface charge densities at both interfaces is computed to be approximately $-0.015$ e/nm$^{2}$ in agreement with electrophoretic experiments. We also show, using an energy decomposition analysis, that the electronic origin of this phenomena is rooted in a collective polarization and charge transfer effect.
\end{abstract}

\section{Introduction}

Air bubbles and oil droplets in water move towards the anode under the
presence of electric fields\cite{zetahunter1988,carruthers1938,takahashi2005} implying that they develop a natural negative charge. This experimental observation has served as
the seed for one of the most hotly debated topics in physical
chemistry for the last several decades with numerous conflicting
intepretations underlying the origins of the negative charge\cite{Saykally2013,wealebeattie2009,beattiefaraday2009,jungwirth2009,Mishra2012}.
Here we make an important leap in our understanding of this
problem providing a framework that helps rationalize the observed
phenomena that is rooted in what we propose is a generic charge transfer mechanism associated with the interfacial structure  at the surface of water and oil.

The electrokinetic experiments of both air and oil droplets paint a
very similar picture despite having rather disparate chemistries
namely, that  the zeta potential ($\zeta$) vanishes under acidic
conditions and that with increasing pH, it becomes  increasingly
negative showing that these interfaces retain a negative charge at
neutral pH\cite{takahashi2005}. If waters constituent ions, the proton and
hydroxide are the only sources of charging in water, these
experiments suggest that the negatively charged OH$^{-}$ ions stick to
the surfaces with binding energies on the order of 10-20 times larger
than thermal energy\cite{beattiefaraday2009,wealebeattie2009}! 

This interpretation has been heavily contested from both experimental
and theoretical fronts. Spectroscopy of interfaces using second harmonic generation (SHG) and sum frequency generation (SFG) tell a less
consistent picture ranging from the presence of H$^{+}$ at the surface
under acidic conditions to weak binding of the OH$^{-}$
under basic conditions\cite{PetersenSaykally2005,petersen2008,Shen2008,tarbuck2006,Shultz2006}. There have also been other suggestions
implicating hydrocarbon impurities such as bi-carbonate ions\cite{Ganachaud2018} as the source of the surface charge
although this seems to have been ruled out in very recent experimental
work interpreting the Jones-Ray effect\cite{okur2018}.  Theory and simulations continue to play an important
role in the interpretation of these experiments. Several groups have
pioneered insightful \emph{ab initio}\cite{Buch2007,Baer2014,mundy2009}, empirical valence bond\cite{voth1,voth2,voth3} and more recently classical empirical potential based\cite{Netz2017} molecular dynamics simulations of the air-water interface. Most of these studies indicate
that the H$^{+}$ has some marginal preferential binding to the surface
of water while the OH$^{-}$ is effectively repelled from this
interface. 

In this report, we assert that the negative charge at extended
hydrophobic interfaces does not require the binding of OH$^{-}$ ions.
Using state-of-the-art linear-scaling density functional theory
(LS-DFT) based simulations of thousands of atoms, we elucidate the
electronic properties of two paradigmatic systems: the air and
oil-water interfaces. Both systems are characterized by a regime of
significant negative charge that is primarily modulated by how charge
transfer changes for different water defects\cite{gasparotto2016}.  Asymmetries in hydrogen bonding patterns between acceptors and donors can cause
rather subtle but sizable swings in the charge of water  molecules. 

At the surface of water, the charge transfer leads to a triple layer of
charge with negative surface charge a couple of Angstroms from the
surface. While a similar effect occurs at the oil-water interface, an
additional complexity emerges: there is some transfer of charge from
the water to the oil molecules leaving the latter negatively charged.
Interestingly, we show using an energy decomposition analysis that despite the low-dielectric character of the oily molecules, those at the
surface experience subtle electronic effects involving both polarization and charge transfer.
The surface charge densities that we determine are an order of magnitude larger than those discovered in previous studies\cite{jungwirth2012,wickrick2012,samson2014} bringing them in closer agreement with experiments.

Besides the direct implications in providing fundamentally new insights into
the controversy, our results have potential for broader
impact. The behavior of hydrophobic interfaces such as the ones we
have tackled here, also lie at the heart of fundamental questions in
atmospheric\cite{hendrik2016} and pre-biotic chemistry\cite{Griffith2012,Mompean2019} where catalysis plays a critical role. In addition to electrophoretic experiments, there is
also a growing literature in the area of triboelectrification, where
water near certain hydrophobic interfaces is observed to receive
negative charge\cite{Lin2013} or even a more patchy surface charge distribution at the boundary between different dielectric media\cite{Baytekin2011}. The electronic and molecular origins of the
underlying phenomena in these contexts, remains poorly understood. The
charge transfer mechanisms espoused in this contribution will help broaden
the scope of the discussion in the area of contact electrification.

\section{Computational Methods}

Most of our results rely on simulations of two different systems that
serve as prototypical models for water near hydrophobic surfaces: the
surface of water and water near oil. The air-water interface was
modeled building a water slab with dimension 40\AA\ X 40\AA\ X
40\AA\ and adding a vacuum padding of 80\AA\ on each side of the slab
along the z direction. This system comprises of a total of 6540 atoms
(2180 water molecules). The second system studied, was a water-oil
interface where the oil phase was composed of 200 dodecane molecules,
while the water phase was made up of 1960 water molecules leading to a
total of 13479 atoms. The cell dimensions for this system are
46.269\AA\ X 46.269\AA\ X 62.768\AA. 

Our strategy for performing the analysis involved two steps: firstly,
classical empirical molecular dynamics simulations were performed in
order to allow for large and long-time scale fluctuations that would
not be possible using \emph{ab initio} molecular dynamics; secondly,
configurations from these simulations were sampled from which the
electronic structure calculations were performed. The classical MD
simulations were run using the GROMACS software\cite{GROMA}. The water
phase was modeled using the TIP4P/2005\cite{TIP-pap} force field in both cases. The
dodecane molecules were simulated using the modified OPLS-AA (L-OPLS)
potential developed by B{\"o}ckmann et al.\cite{OIL-FF}. This
potential has been parametrised on the basis of high level ab-initio
calculations, densities and heats of vaporization of both short- and
long-chain alkanes and on the phase transition temperature of
pentadecane in order to extend the OPLS-AA validity to long
hydrocarbon chains and recover a more precise description of their
phase transition temperatures and ordering.

The air-water system was equilibrated for 10 ns using the NPT ensemble
using the Parrinello-Rahaman barostat \cite{Barost} for the first half of the run for bulk water, 
followed thereafter by an NVT simulation at 300K opening up a gap in the z-direction which separated the two water surfaces by 160\AA\ for the remainder of the simulation. The production simulations were run for 20 ns. The surface tension computed from the classical simulations is 68.2 in agreement
with previous studies\cite{SurfTens}. From these simulations a total of 250
configurations were randomly selected to perform the electronic
structure calculations. The oil-water interfaces were equilibrated first
20 ns via NVT simulations. The production calculations were then run
for 40 ns using the isothermal-isobaric ensemble. 200 frames were
randomly selected for the electronic structure calculations.

In order to extend the scope of our  electronic structure
calculations, we employed a Linear Scaling DFT (LS-DFT) approach as
implemented in the ONETEP code\cite{ONET}.  This technique allowed us
to extend the system sizes in our study to thousands of atoms and to
model electronic effects of extended hydrophobic interfaces on the
nanometer lengthscale. For a more detailed and technical summary of
the underlying theory the  reader is referred to relevant
literature\cite{LS-Rev}. Here we briefly summarize the essential
ideas. LS-DFT as implemented in ONETEP makes use of the
\emph{nearsightedness}\cite{Nearsight} inherent to quantum many-body
systems by exploiting the single-particle density matrix,
$\rho$(\textbf{r}, \textbf{r'}) \cite{DM1,DM2} representation of the
system of interest. Within ONETEP, $\rho$(\textbf{r}, \textbf{r'}), is
expressed in a separable form \cite{DM3,DM4} via atom-centered
functions (non-orthogonal generalized Wannier functions, NWGFs
\cite{NGWF}), $\phi_\alpha(\textbf{r})$,  as:\\
\begin{gather}
\rho(\textbf{r},\textbf{r'})=\sum_{\alpha\beta}\phi_\alpha(\textbf{r})\textbf{K}_{\alpha\beta}
\phi^*_\beta(\textbf{r'})
\end{gather}

\noindent

In the above, $\textbf{K}_{\alpha\beta}$ are the matrix elements of the density
kernel, which are nonzero only if
$|\textbf{r}_\alpha-\textbf{r}_\beta| < \textbf{r}_c$, with
$\textbf{r}_\alpha$ and $\textbf{r}_\beta$ representing the
coordinates of the centers of $\phi_\alpha$ and $\phi_\beta$, and
$\textbf{r}_c$ is a real-space cut-off threshold.  The truncation of
the density kernel ($\textbf{K}_{\alpha\beta}$) is validated by the
exponential decay of $\rho(\textbf{r},\textbf{r'})$ with respect to
$|\textbf{r}-\textbf{r'}|$ for systems with an electronic band gap
\cite{DM-BG}. Such truncation leads to a sparse density matrix
($\rho(\textbf{r},\textbf{r'})$) that makes any insulating or
semi-conducting systems (including the different interfaces considered
here) treatable using linear scaling simulation. The NGWFs are centered
on the nuclear coordinates and localized within a sphere of radius
$r_\alpha$. Their non-orthogonality, implies a non-diagonal overlap
matrix, $S_{\alpha\beta}$:

\begin{gather}
\textbf{S}\alpha\beta=\int d\textbf{r}\phi^*_\alpha(\textbf{r})\phi_\beta(\textbf{r})
\end{gather}

\noindent

In practice the NGWFs are expressed as a linear combination of
coefficients $C_{m\alpha}$, of localized but orthogonal periodic
cardinal sine (psinc) functions \cite{NGWF}, $D_m(\textbf{r})$, as:

\begin{gather}
\phi_\alpha=\sum_{m}C_{m\alpha}D_m(\textbf{r}-\textbf{r}_m)
\end{gather}

\noindent
with m indexing the real-space Cartesian grid inside the spherical
localization region of $\phi_{\alpha}$. The psinc functions are
obtained from a discrete sum of plane-waves, that makes the set of
$D_m(\textbf{r})$ independent of the nuclear coordinates and
systematically improvable upon increase of the kinetic energy cutoff
\cite{NGWF}. The convergence of the ONETEP approach is then dependent
on interlinked computational factors such as the kinetic energy
cutoff, the number of NGWFs ($\phi_\alpha$) per atom and their
localization radius.

In our LS-DFT calculations, the adopted kinetic energy cutoff was 1000
eV and 4 NGWFs were used for O atoms and 1 NGWF was used for the H
atoms. In all cases, no truncation of the density kernel
($K_{\alpha\beta}$) was enforced. The localization radius for the
NGWFs was 10 Bohr in all cases. These parameters were chosen after a
careful benchmark for the water monomer and dimer porperties against
the ab-initio code CP2K\cite{cp2kqui}. Simulations were performed using the
BLYP\cite{B88pap, LYPpap} functional with Grimme’s D2\cite{GD2} empirical dispersion
corrections. In all cases, separable (Kleinman-Bylander)\cite{KLpseudo}
norm-conserving pseudopotentials constructed with the Opium
code\cite{opiumsite} were used. 

In order to characterize any possible charge-gradients developing
at our simulated interfaces , atomic point charges are then derived 
from the electron density. The atomic charges reported in this work were calculated
using the DDEC3\cite{DDECbas} scheme implementation in ONETEP \cite{ONEDDEC}.
DDEC3 is an atoms in molecules (AIM) approach where the total 
QM electron density ($n(r)$)
is partitioned into overlapping atomic densities ($n_i(r)$):

\begin{gather}
n_{i}(r)=\frac{w_{i}(r)}{\sum_{k}w_{k}(r)}n(r)
\end{gather}

The atomic partial charges are then computed by integrating the atomic
electron densities over all space:

\begin{gather}
q_i=z_i-N_i=Z_i-\sum{n_i(r)d^3r}
\end{gather}

where $N_i$ is the number of electrons assigned to atom i and $z_i$ is
its effective nuclear charge. In the same fashion, higher-order atomic
multipoles may be computed as first-order, second-order,(etc) moments
of the atomic electron densities. Various definitions of the weighting
factors $w_i(r)$ exist. In the DDEC case the weighting function is
described so that the atomic weights are simultaneously optimized to
resemble the spherical average of $n_i(r)$ and the density of a
reference ion of the same element with the same atomic population
$N_i$. In this way, the assigned atomic densities yield a rapidly
converging multipole expansion of the QM electrostatic potential and
the computed populations are chemically reasonable. A more detailed
description of the method can be found in the following references
\cite{ONEDDEC,DDECbas}.

An important part of our findings reported in this work is the charge
transfer occurring between different types of water molecules at the
surface of water and between water and dodecane at the oil-water
interface. In particular, the sensitivity of our results to choice of
using a standard GGA functional like BLYP was assessed and validated
using both hybrid functionals such as B3LYP\cite{B3LYPpap} as well as wavefunction approaches such as MP2\cite{MP2pap}. In similar spirit, the charge transfer
between water and decane is also found to be generic result with both B3LYP and MP2. Besides the quality of the electronic structure,  we also examined the
sensitivity of the charge transfer to sampling configurations of the air-water
interface sampled from mb-POL\cite{MBPOLpap} as well using different charge-paritioning schemes. Details of these benchmarks are elucidated in the SI. All in all, we demonstrate that our findings are on firm theoretical footing.

\pagebreak
\section{Results and discussion}

In the ensuing analysis, we begin by first discussing the charge
gradients observed at the surface of water and at the oil-water
interface. We subsequently compare and contrast the microscopic
origins of the charge oscillations in the two systems.

\subsection{The Air/Oil Water Interfaces are Negatively Charged}

\begin{figure*}[htb]
\begin{center}
\includegraphics[clip=true, width=0.83 \textwidth]{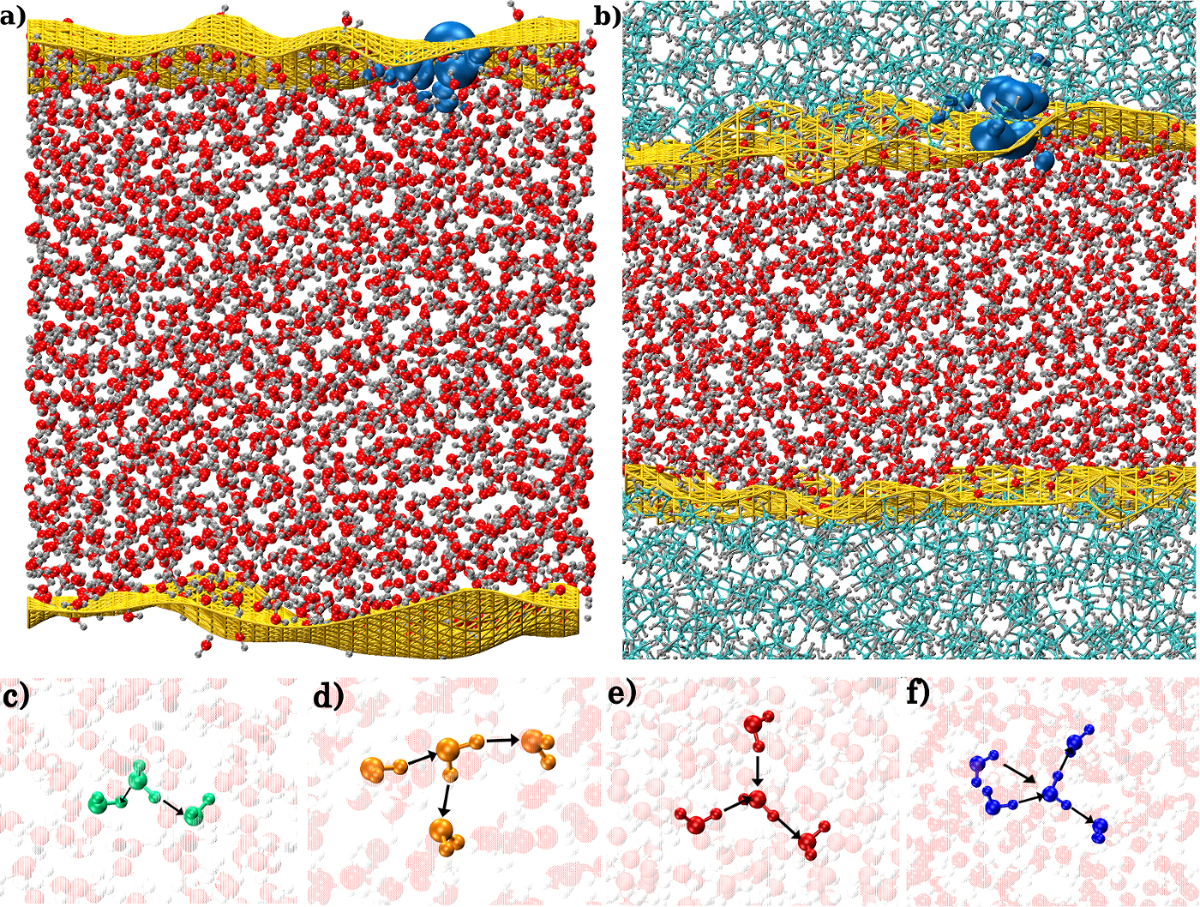}
\caption{Outline of the two simulated systems: a) the air-water and b) the oil-water. The Willard-Chandler instantaneous surfaces are highlighted in yellow. The Highest Occupied Molecular Orbital (HOMO) in both systems is highlighted in cyan. Panels c), d), e), f) represent the different types of water molecules coordination configurations that will form the focus of our discussion later. For each 
coordination configuration the accepted hydrogen bond(s) are represented by an arrow pointing in, while the donated hydrogen bond(s) are shown by an arrow pointing out. For brevity, the following nomenclature will be used in the rest of the paper: donated hydrogen bonds will be called \emph{out} while accepted hydrogen bonds will be called {in}. In this way panels  c), d), e), f) represent respectively 1in-1out, 1in-2out, 2in-1out and 2in-2out water coordination configurations.}
\label{Water-air-outline}
\end{center}
\end{figure*}

The top left and right panels of Figure \ref{Water-air-outline} show snapshots of our two simulated systems: a) the air-water and b) the oil-water systems consisting
of a total of 6540 and 13480 atoms respectively. The yellow-colored surface
corresponds to the Willard-Chandler\cite{Istant-surf} instantaneous interface (WCI) that
is constructed for the water phase. In both figures, the cyan-colored 
isosurface corresponds to the highest molecular orbital which is localized at the interface. To the best
of our knowledge, our simulations represent the first of their kind where
the electronic structure of thousands of atoms of the air and oil-water interfaces are treated.

\begin{figure*}[htb]
\begin{center}
\includegraphics[clip=true, width=0.90 \textwidth]{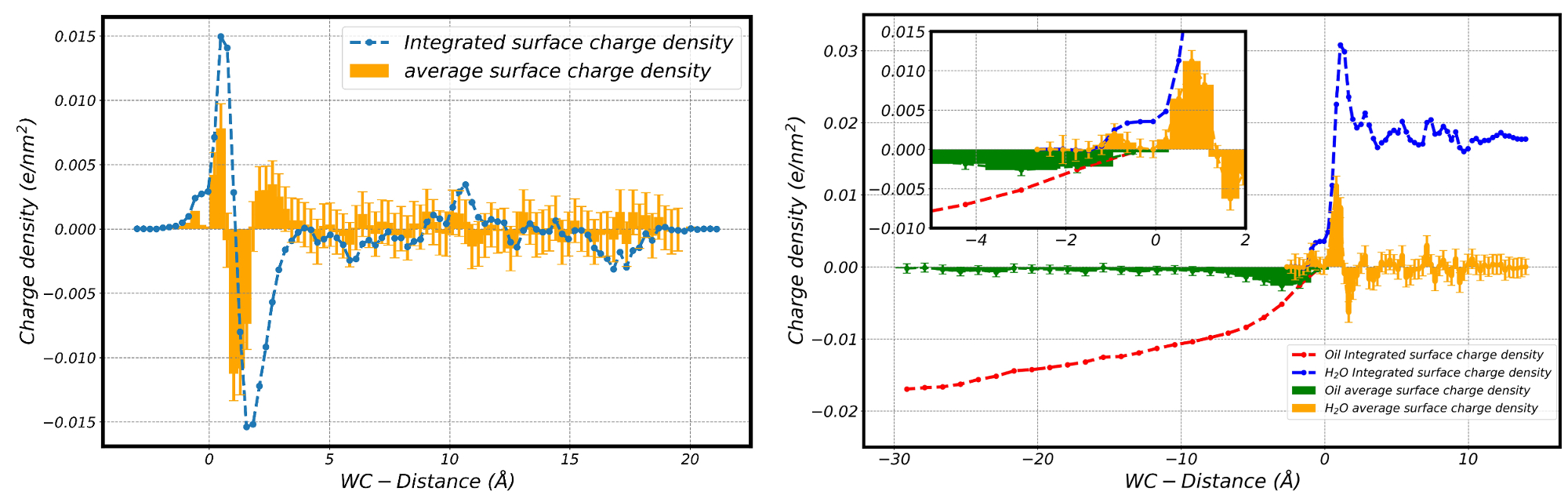}
\caption{Charge density (yellow bars) and integrated surface charge density (blue line) obtained for the DDEC charges extracted from our calculations  for  the  air-water (left panel) and  oil-water (right panel) interfaces. For the oil phase in the right panel, the charge density is reported using green bars and the integrated surface charge density is shown by the red line. The inset in the right panel magnifies the charge oscillations right at the interface between the two phases.}
\label{Water-oil-charge}
\end{center}
\end{figure*}

The possibility that a charge transfer mechanism could rationalize the
negative charge at hydrophobic surfaces has been suggested in previous
theoretical studies\cite{jungwirth2009,jungwirth2012,wickrick2012} although the reported charge densities were too small compared to those obtained from electrophoretic experiments. The essential idea is that asymmetries in hydrogen
bonding between water molecules at an interface, leads to a subsequent
imbalance in the charge transfer along donating versus accepting
hydrogen bonds. We begin by showing in Figure \ref{Water-oil-charge}
the charge densities and integrated surface obtained for the DDEC
charges extracted from our calculations for the air-water and oil-water
interfaces. As alluded to in Figure \ref{Water-air-outline}, in order to perform
this analysis, a description of the corrugations of the interface is
needed. We used a formulation proposed by Willard and Chandler\cite{Istant-surf}
which characterizes the instantaneous density fluctuations of the interface.
The zero on the x-axes of Figure \ref{Water-oil-charge} corresponds to
the position of the WCI.

The left panel of Figure \ref{Water-oil-charge} reveals the  presence
of significant charge gradients at the air-water interface covering a
length scale of about 5 \AA. In particular, we observe a triple layer
of charge (green bars): above the instantaneous interface there is a
positively charged layer (1st layer) of thickness 1.8\AA\ with a
charge density of $\sim 0.0075\ e/nm^3$; immediately below
this, there is a compensating negative layer of similar thickness
(2nd layer) but with a larger charge density of $\sim-0.012\ e/nm^3$; 
and finally, below 2-5\AA \ from the interface, there
is another positively charged layer (3rd layer, charge density
$\sim 0.003\ e/nm^3$) after which, charge neutrality develops (the 4th layer). 
The presence of the triple layer is essentially caused by the asymmetry in
the magnitude of the first two charged layers where the negative
branch is about twice as large than that of its positive counterpart closer to
the interface. The charge density shown, can be integrated from the
vacuum to the bulk to give a better description of the cumulative surface
charge. The dashed blue curve shows the integrated surface charge
density - between 1-2\AA \ from the WCI interface there is a
substantial negative charge density of $\sim -0.015\ e/nm^2$ which is
about an order of magnitude larger than previous findings\cite{jungwirth2012,wickrick2012}.

How does the behavior change near the oil-water interface? Rather
unexpectedly, the oil phase is not a passive spectator in the charging
mechanism.  The right panel of Figure \ref{Water-oil-charge} shows
similar distributions in charge density and integrated surface charge
for both the water \emph{and} the oil.  As before, the zero
on the x-axis corresponds to the position of the WCI
surface. In stark contrast to the surface of water, there is
a sharp peak of positive charge in the water phase just\ \emph{below}
the interface with a charge density of $\sim 0.07 e/nm^3$. This
layer is followed by a negatively charge region of water with a lower
charge density of $\sim$ 0.05 e/nm$^3$. The complementary charge distributions 
of the dodecane molecules forms one of our central findings in this
report namely that the oil phase is negatively charged. 
We observe a large negative surface charge
density of about $-0.012$ e/nm$^{2}$ in the oil phase. These results are striking since
they show that there is a net charge transfer of $\sim$0.4 electron charge
from the water to the oil (Figure 1 SI). Furthermore, the magnitude of this surface charge is very similar to what is observed at the air-water interface, which is also consistent with the similarity in the zeta-potentials obtained from air and oil droplets\cite{agmon2016}.

\subsection{Charging is Coupled to the Local Topology and Environment}

In order to dissect the microscopic origins of the charge gradients
observed at the interface, we turn next to examining how the charge is
modulated by differences in topological defects.  It is well appreciated,
that fluctuations in liquid water both in the bulk\cite{gasparotto2016} and at
surface\cite{gibertihassanali2017}, create local coordination defects which have
asymmetries in the number of donated versus accepted hydrogen
bonds. The bottom panel of Figure \ref{Water-air-outline} depicts different types of water molecules that will form the focus of our discussion later ranging
from the canonical tetrahedral waters that accept and donate two
hydrogen bonds (2in-2out) to various other undercoordinated defects such as
those that accept one/two and donate two/one hydrogen bonds (1in-2out and 
2in-1out waters as described in the caption).

Using the four layers previously defined to describe the different
charged layers, we determined how the relative concentration of
different water molecules change as one moves from the interface to
the bulk, as well as their relative contribution to the total charge.
For clarity, the contributions to the total charge are given
separately for the species that are positive or negative.  In all the
layers we show only the most dominant coordination states: for the
first layer, this involves all species with a total population $>$
0.5\%, while for the other layers it is those that contribute at least
2\% of the total population.

The 1st layer is dominated by many undercoordinated species due to the presence of the dangling O-H bonds\cite{mischabonn2011}. Particularly relevant is the role of the 1in-0out and 2in-1out species both of which lead to a net positive contribution of charge within the first layer.  It is also striking to note that in spite of being balanced in terms of hydrogen bonds, the 1in-1out species noticeably contributes negatively to the overall charge underlining the importance that the asymmetry of hydrogen bonding itself does not exclusively control the charging behavior and that there are important collective polarization effects.
 
\begin{figure*}[htb]
\begin{center}
\includegraphics[clip=true, width=0.75\textwidth]{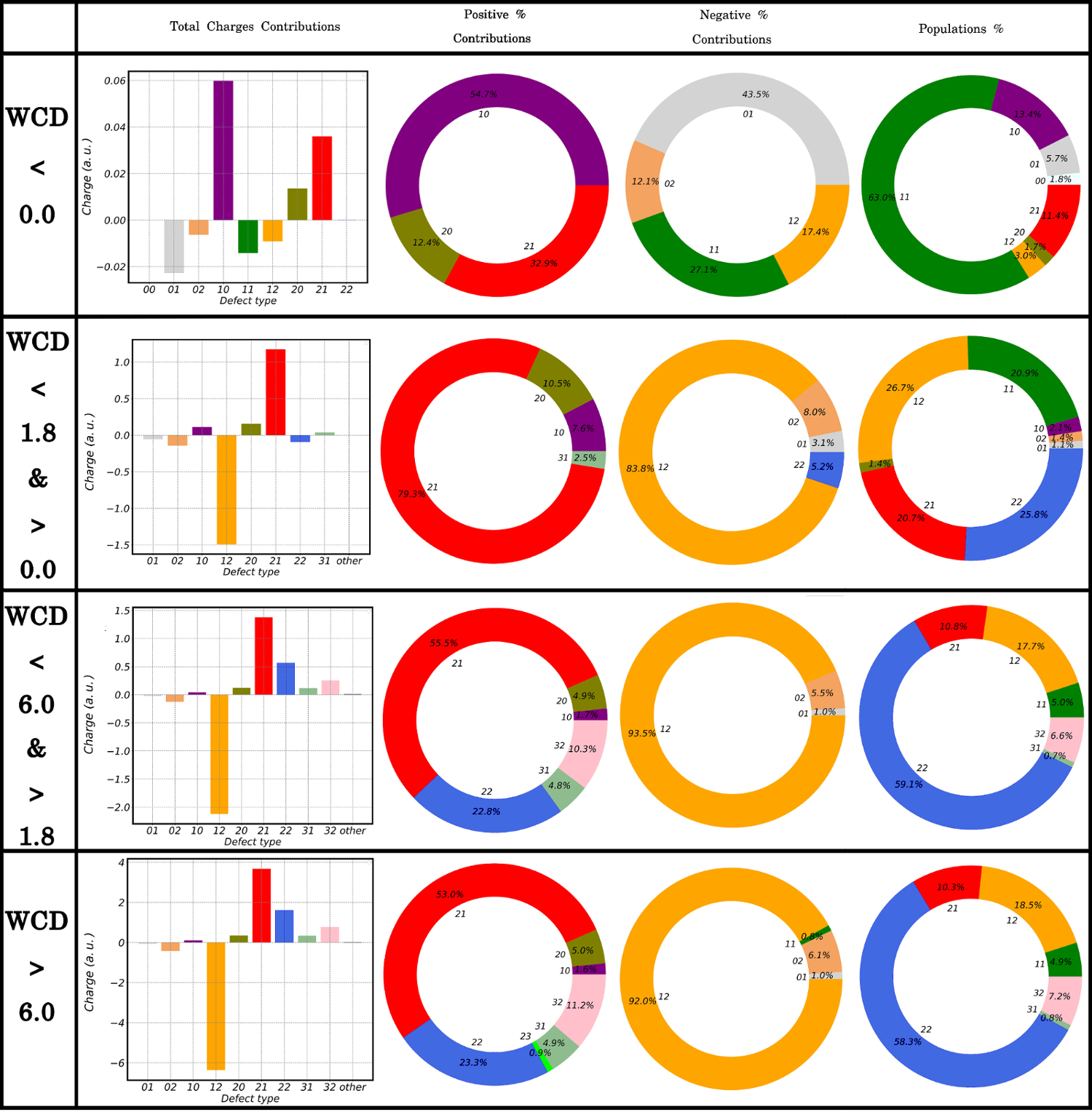}
\caption{Analysis of the water-air interface layer-by-layer (as defined in the main text) in terms of the average total charge contribution, percentage contribution to the positive charge, percentage contribution to the negative charge and percentage contribution to the overall population for each water coordination configuration.}
\label{Water-air-pop}
\end{center}
\end{figure*}

Layer two presents some drastic differences with respect to first and
it is this region that provides important clues into the origin of the
negative charge at the surface of water. The negative oscillation in
charge between 0 and 1.8\AA\ below the WCI is dominated by the
competition between the charging behavior of the 1in-2out and 2in-1out
water molecules (second uppermost panel Figure \ref{Water-air-pop}). Contrary to what one might have expected, the behavior is not symmetric. The total absolute
charge of the 1in-2out waters is larger than 2in-1out by $\sim$0.3
e (corresponding to a surface charge density of $-0.019\ e/nm^3$
). Interestingly, this behavior is not exclusive to the interface
but is a feature that continues to occur even in the bulk. Part of this effect 
originates from the fact that the concentration of 1in-2out waters is
larger than 2in-1out waters consistent with previous studies\cite{gasparotto2016}  although the role of this difference in the creation of charge gradients has not been recognized until this point. Besides this, we will see later that charging behavior of the these two types of water defects are not symmetric across the region of the interface.

Despite its relatively lower weight, it is interesting to underline
the negative contribution that the tetrahedral waters give to the
total charge in the second layer. This behavior then swings in the
opposite direction in the 3rd and 4th layers where the 2in-2out waters 
integrate to a net small positive charge. Similar to the
traits of the 1in-1out water molecules right at the top of
the surface of water, we find that the charge fluctuations are
affected by other factors beyond the asymmetry in hydrogen bonding.
In addition, what is also particularly important to observe
as we move across different layers is that
the dominant positive-negative branches of the 1in-2out and 2in-1out
waters occur even in the bulk phase. Charge neutrality in bulk liquid
water involves a complex mix of the 2in-2out, 3in-1out, 3in-2out and
2in-0out essentially counterbalancing the negative contribution
of the 1in-2out water molecules that is not achieved by the 2in-1out
defects.

\begin{figure*}[htb]
\begin{center}
\includegraphics[clip=true,width=0.80\textwidth]{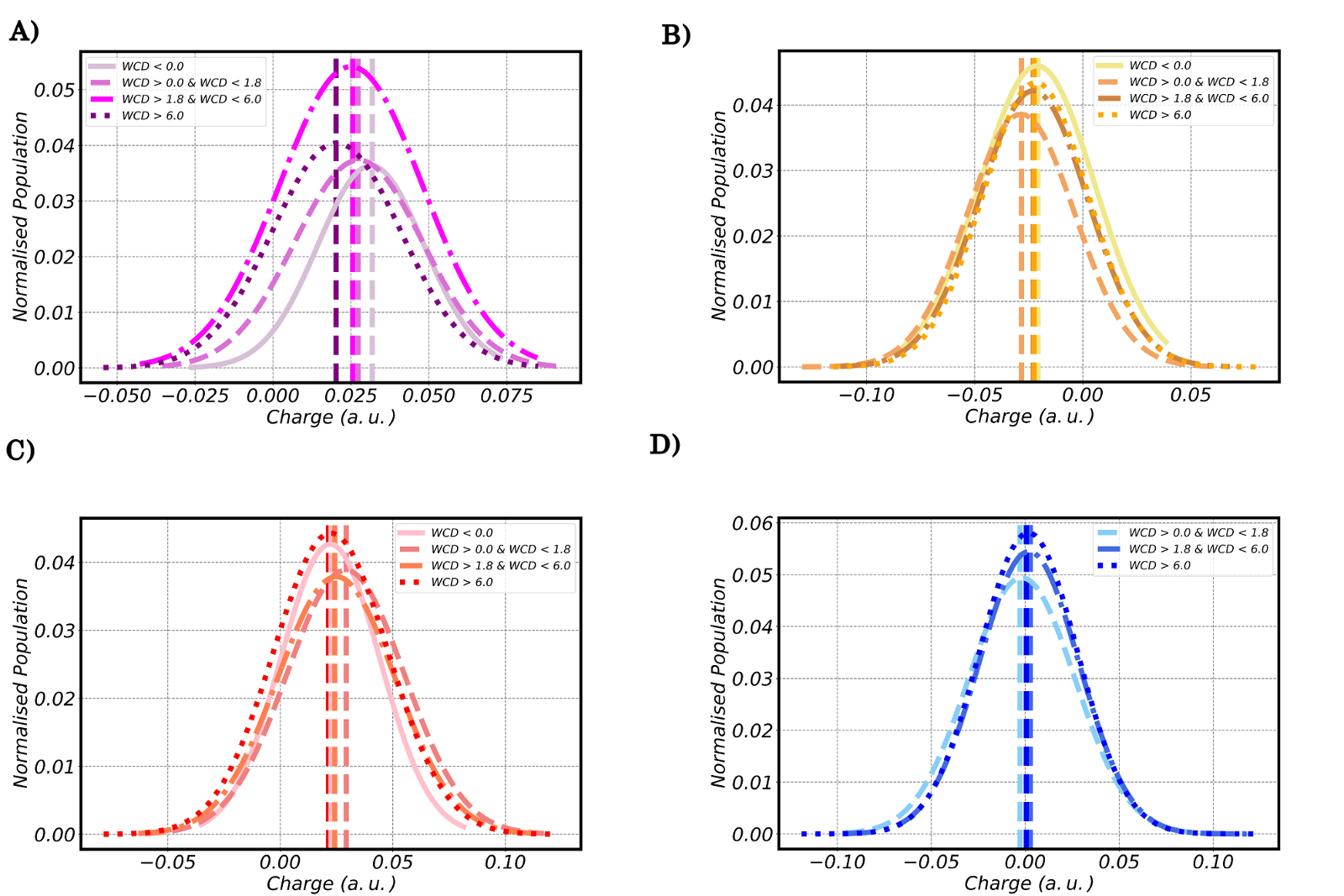}
\caption{ Layer-by-layer charge distributions of the 1in-out (panel A), 1in-2out (B), 2in-1out (C) and 2in-2out (D) water molecules for the water-air interface. The first moment is highlighted by a vertical line for each distribution.}
\label{Water-air-dist}
\end{center}
\end{figure*}

Figure \ref{Water-air-pop} provides a collective picture of the role of water-topology on charging but does not tell us anything about the nature of the
fluctuations of individual molecules and how they are perturbed by the
interface. In order to explicate this, we show in Figure
\ref{Water-air-dist} the charge distributions of the 2in-1out,
1in-2out, 2in-2out and 1in-0out water molecules in the four
layers. These distributions confirm our intuitions built on the
preceding analysis that the interfacial region agitates water
molecules in subtle and very surprising ways. There is clearly a
change in the average charge for water molecules in different
layers. A more quantitative analysis of the differences can also be
obtained by the first four moments reported in Figure 2 of the Supporting Information. These results show that the for the water-air interface the average charge on each water molecule for the 1in-2out, 2in-1out species increase form the 1st to the 2nd layer (right at the negative oscillation) just to decay again in the bulk like region. An increment of the average charge for the 1in-0out molecules is also observed at the surface. This increment, however, decrease monotonically moving into the bulk affirming the important role that this species has mainly for surface properties. As previously reported the tetrahedral coordinated waters in the 2nd layer are on average negatively charged and then regain a positive character towards the bulk. While the average charge changes between the 2nd and 3rd layer are moderate ($\sim$0.005 e on average) the huge increment in weight of the 2in-2out species between these two layers lead to sizeable changes in the overall charge contributions. This observation further solidifies the assumption that small variation in the average charge per molecules can lead to big changes in the interfacial system behaviour. These types of features would not be captured by models using a constant charge transfer value\cite{jungwirth2012,wickrick2012} that is not modulated by the local environment.

\begin{figure*}[htb]
\begin{center}
\includegraphics[clip=true, width=0.75\textwidth]{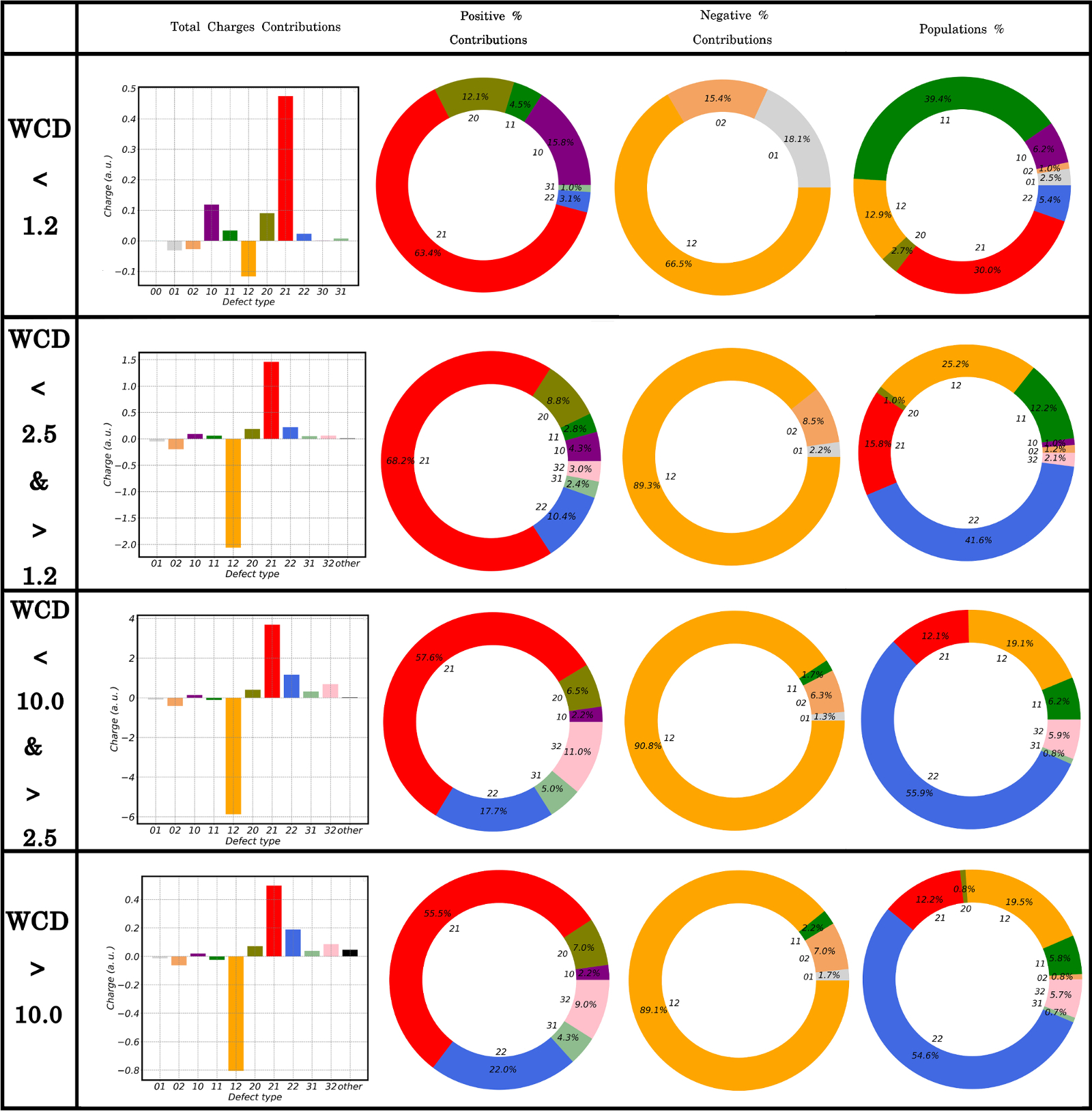}
\caption{Analysis of the water-oil interface layer-by-layer (as defined in the main text) in terms of the average total charge contribution, percentage contribution to the positive charge, percentage contribution to the negative charge and percentage contribution to the overall population for each water coordination configuration.}
\label{Water-oil-pop}
\end{center}
\end{figure*}

\begin{figure*}[htb]
\begin{center}
\includegraphics[clip=true, width=0.80\textwidth]{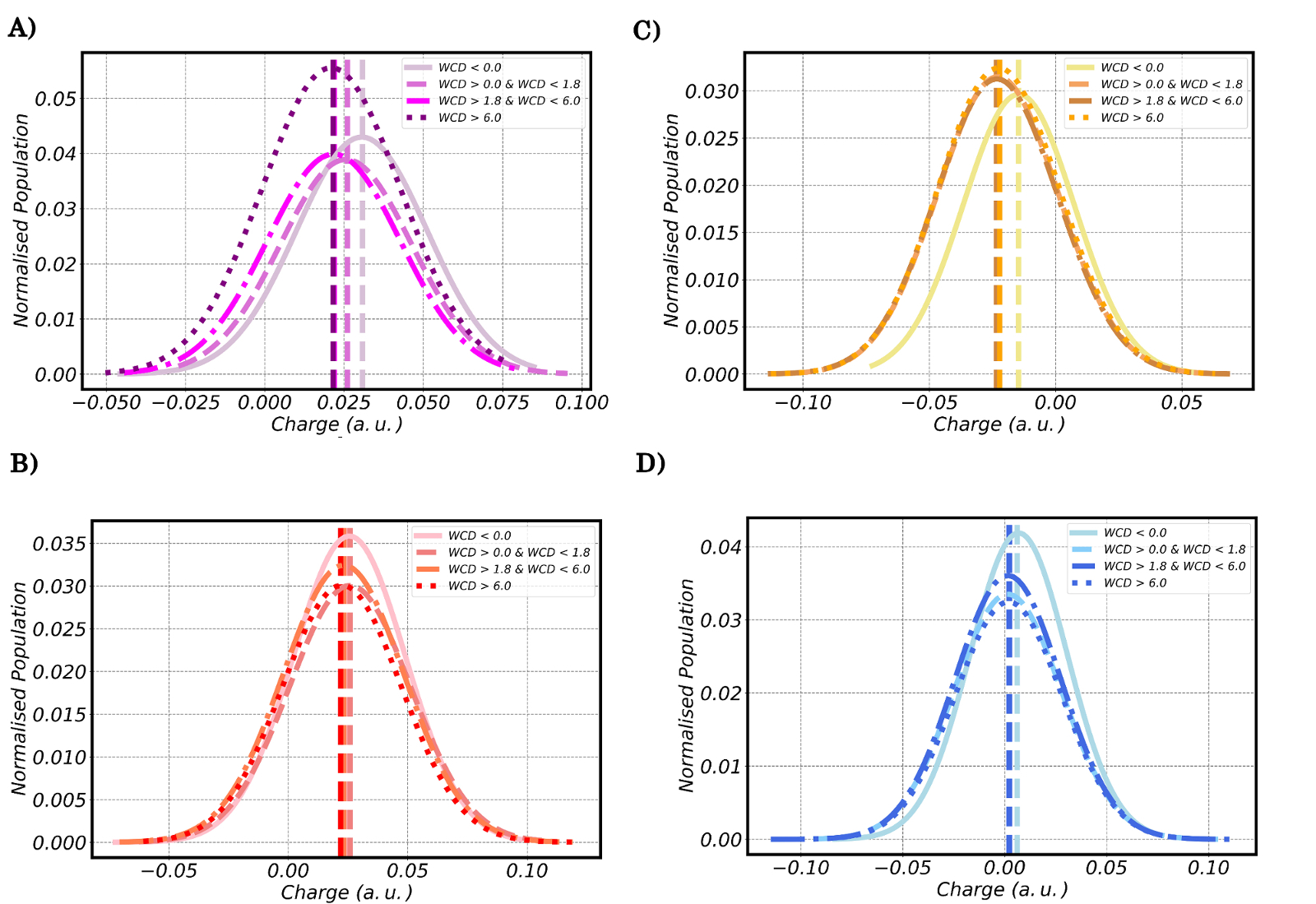}
\caption{Layer-by-layer charge distributions of the 1in-out (panel A), 1in-2out (B), 2in-1out (C) and 2in-2out (D) water molecules for the water-oil interface. The first moment is highlighted by a vertical line for each distribution.}
\label{Water-oil-dist}
\end{center}
\end{figure*}

Having understood the origins of the charge oscillations at the
surface of water, we move next to examining how these properties
behave near oil. In order to adequately compare the two interfaces we
repeated the charge and population analysis previously done for the
water-air system as a function of different layers (Figure \ref{Water-oil-pop})  for
the oil-water interface. The interfacial structure of water near the
oil surface is quite different from that at the  surface of water. The
1st layer is characterized by an increased presence of the 2in-1out
coordination defects. In addition, the 1in-1out water molecules
contribute positively to the overall charge. These two effects
conspire together to produce a much larger positive charge in the
first layer compared to the surface of water. This is in essence
activated by the transfer of charge to the oil phase. The 2nd layer
derives very similar trends with respect to its air-water
equivalent, except for the fact that in this case, the tetrahedral
2in-2out water integrate to a net positive charge. Beyond the 2nd
layer, the behavior is very similar to that observed in Figure \ref{Water-air-pop}.
Although the individual charge distributions for water near
the oil show some differences (\ref{Water-oil-dist} and Figure 2 SI, lower Table), the overall behavior is very similar -specifically, charge fluctuations of different water topologies
and their sensitivity to proximity to the interface appears to
be an important part of the physics of charging.

Figures \ref{Water-air-dist} and \ref{Water-oil-dist} for the air and oil-water interfaces show that the charge on different water molecules sustains rather large fluctuations. Furthermore, the fact that tetrahedral waters can conspire to produce a positive total charge, suggests a highly non-trivial asymmetry in the charge transfer occurring on the donating versus accepting side of the hydrogen bond network. Part of the origin of these asymmetries lie in the differences in the manner in which charge on water molecules is the fluctuations in the local tetrahedrality as seen in previous studies\cite{Kuhne2013}.

\subsection{The Role of Electronic Polarization and Charge Transfer}

The stabilizing role of charge transfer between the hydrogen bonds of water molecules has been well appreciated in the literature\cite{bellkhaliullin2009,Kuhne2013}. Its role however, at the surface of water and how it is modulated by defect fluctuations has not been recognized until this point. The situation near the oil-water interface is much more surprising and warrants a deeper examination.

\begin{figure*}[htb]
\begin{center}
\includegraphics[clip=true, width=0.95\textwidth]{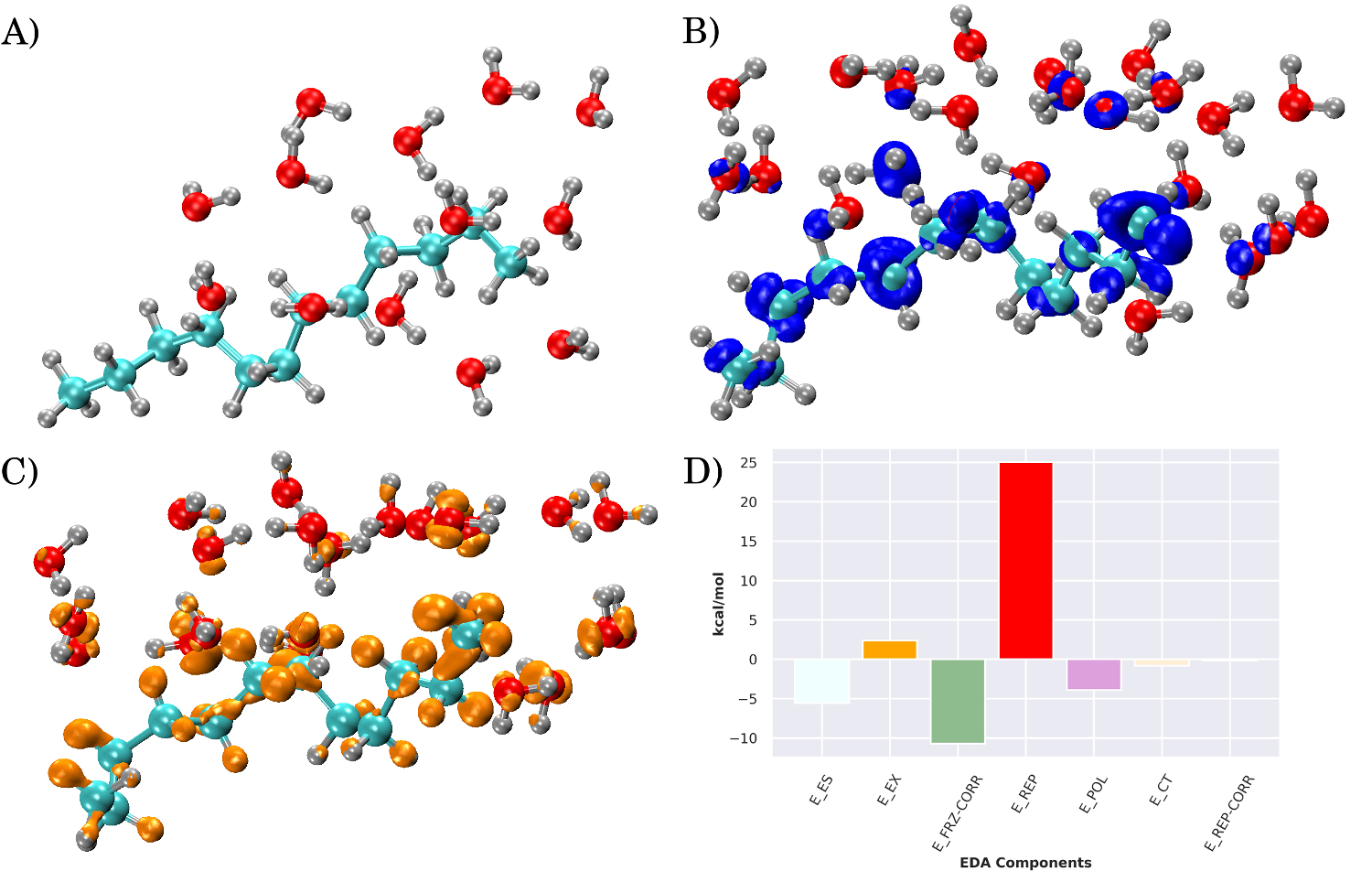}
\caption{ Sample representation of the clusters considered for the EDA calculations (Panel A). The Polarization (blue) and Charge Transfer (orange) Electron Density Difference for the same sample oil-water cluster are highlighted respectively in Panel B and C. The average energy component contributions calculated via EDA analysis for the 60 oil-water clusters are reported in Panel D. The acronyms of the different components are: E$_{ES}$ is the Electrostatic Energy, E$_{EX}$ is the Frozen Exchange Energy, E$_{FRZ-CORR}$ is the Frozen Correlation, E$_{REP}$ is the Pauli-Repulsion Energy, E$_{POL}$ is the (Stoll SCF-MI) Polarisation Energy, E$_{CT}$ is the Charge Transfer Energy and E$_{REP-CORR}$ is the Repulsion Correlation.}
\label{EDA-EDD}
\end{center}
\end{figure*}

In order to understand better the underlying quantum mechanical effects associated with the build up of surface charge, we performed an Energy Decomposition Analysis (EDA) as implemented in ONETEP\cite{EDA-ON}.
The EDA analysis essentially provides a framework to disentangle various contributions of the interaction energy between the water and decane coming from electrostatics, polarization, exchange and charge transfer. 
The EDA analysis reported here was performed on approximately 60 clusters consisting of one decane molecule and all water molecules within 3.5 \AA from it. The clusters were carved out from the thermal simulations described earlier and typically consist of about 12 H$_2$O molecules (see Figure \ref{EDA-EDD}A).
For this subset of clusters, the average DDEC charge on the decane molecules was -0.056 e.

We begin by first showing the qualitative behavior involving the intra and inter-fragment electron reorganization in the clusters. Panels B and C of Figure \ref{EDA-EDD} shows the electron density difference (EDD) surfaces that are obtained from the EDA calculations between intermediate states involving the extraction of the polarization (blue iso-surface) and charge transfer (yellow iso-surface) contributions (see Methods for more details). Interestingly, we observe that the polarization EDD mostly involves reorganization along the backbone carbon atoms of the decane. On the other hand, charge transfer appears as response of the electron density that is mainly localized on the hydrogen atoms of decane. In both cases, nearby water molecules also exhibit perturbations from both effects.

A more quantitative measure of these effects at the oil-water interface can be explicated by examining the distributions associated with the various components of the interaction energy. Panel D of Figure \ref{EDA-EDD} shows those components that are repulsive and attractive. The former are dominated by Pauli repulsion and exchange effects. On the other hand, charge transfer and polarization energetics contributes an attractive interaction that adds up to about 5 kcal/mol - for the clusters, this implies a binding energy of slightly under thermal energy per water molecule. As result of these charge transfer and polarization effects that render the oil phase negative and the water one positive the electrostatic component of the Energy is also attractive ($\sim$ 5 kcal/mol). Our findings in the previous section strongly suggest that these effects are likely to be enhanced and also critical driving force in enhancing the negative surface charge observed at extended oil-water interfaces.

\subsection{Discussion and Conclusions}

The findings of our work give strong indication that different types of charge-transfer mechanisms can lead to a significant build up of surface charge density at hydrophobic interfaces. One of the essential ingredients associated with this phenomena, is the presence of local topological defects involving under-coordinated water molecules at the interface. These types of waters are most susceptible to asymmetries in the magnitude of charge transfer between the accepting and donating side which can leave them with a slight bias to take on either a positive or negative charge.

Near the surface of oil, something quite surprising happens: water defects can now inject some electron density into the oil phase leaving the latter negatively charged. The effective surface charge densities derived at the surface of oil is very similar to that near the air-water interface and is also in very good agreement with measurements from electrophoretic experiments. Lurking behind the link between the simulations and these experiments is the question of where the boundary of the slip plane should be placed in Figure \ref{Water-oil-charge}. 

Roke and co-workers proposed a qualitative mechanism (see Figure 5 of Ref. 26) for the effect of charge transfer on electrophoretic mobility. In this scenario, the slip plane resides several Angstroms from the oil surface and the role of fluctuations of water topologies is not considered. For the oil-water interface, our findings suggest that the slip plane will be pinned right near the oil droplet since it is negatively charged while the water in close proximity is positively charged. On the other hand for the air-bubbles, the slip plane must reside somewhere in between the boundary of the second and third charged layers (shown in the left panel of Figure \ref{Water-oil-charge}).

A set of experiments that may also deserve another look at are surface sensitive photoelectron spectroscopy. Winter and Jungwirth combined both theory and experiments in a series of papers to show that the valence band edge of sodium hydroxide solution could be used to interrogate the presence of hydroxide ions at the surface of water\cite{winterjungwirth2009}. They found that the presence of the hydroxide ion could be identified by an enhancement in the states close to the valence band. Although it is beyond the scope of the current report, we examined the projected density of states (PDOS) on water molecules residing in the negatively charged layer at the surface of water and found that they contribute substantially to the valence band (see SI Figure 3,4 and 5). Qualitatively, the defects at the surface are about 0.5eV higher in energy than the mostly tetrahedral ones in the bulk (see SI Figure 3 upper-left panel) - this is indicates an uncanny similarity between the anionic defect, the hydroxide ion, and defects in neutral water.

The mechanism associated with the response of the electronic degrees of freedom, as revealed by an energy decomposition analysis, involve a combination of both polarization and charge transfer. The coupling between the electronic reorganization and nuclear coordinates such as local topology, opens up some very interesting perspectives and questions for future lines of research. An obvious one centers around the generality of our results to other extended hydrophobic interfaces or more heterogeneous surfaces such as metal-oxides or even biological systems. In addition, the patchiness of surface charge at different interfaces deserves further investigation\cite{Baytekin2011}.

\begin{acknowledgement}
EP and AH acknowledge the support from KAUST (Shaheen II project-k1263). EP and AH acknowledge Himashu Mishra for the useful discussions and insightful help provided during the work. 

\end{acknowledgement}

\begin{suppinfo}

\end{suppinfo}

\bibliography{ms}

\providecommand{\latin}[1]{#1}
\makeatletter
\providecommand{\doi}
  {\begingroup\let\do\@makeother\dospecials
  \catcode`\{=1 \catcode`\}=2 \doi@aux}
\providecommand{\doi@aux}[1]{\endgroup\texttt{#1}}
\makeatother
\providecommand*\mcitethebibliography{\thebibliography}
\csname @ifundefined\endcsname{endmcitethebibliography}
  {\let\endmcitethebibliography\endthebibliography}{}
\begin{mcitethebibliography}{65}
\providecommand*\natexlab[1]{#1}
\providecommand*\mciteSetBstSublistMode[1]{}
\providecommand*\mciteSetBstMaxWidthForm[2]{}
\providecommand*\mciteBstWouldAddEndPuncttrue
  {\def\EndOfBibitem{\unskip.}}
\providecommand*\mciteBstWouldAddEndPunctfalse
  {\let\EndOfBibitem\relax}
\providecommand*\mciteSetBstMidEndSepPunct[3]{}
\providecommand*\mciteSetBstSublistLabelBeginEnd[3]{}
\providecommand*\EndOfBibitem{}
\mciteSetBstSublistMode{f}
\mciteSetBstMaxWidthForm{subitem}{(\alph{mcitesubitemcount})}
\mciteSetBstSublistLabelBeginEnd
  {\mcitemaxwidthsubitemform\space}
  {\relax}
  {\relax}

\bibitem[Hunter(1988)]{zetahunter1988}
Hunter,~R.~J. \emph{Zeta Potential in Colloid Science}; Elsevier, 1988\relax
\mciteBstWouldAddEndPuncttrue
\mciteSetBstMidEndSepPunct{\mcitedefaultmidpunct}
{\mcitedefaultendpunct}{\mcitedefaultseppunct}\relax
\EndOfBibitem
\bibitem[Carruthers(1938)]{carruthers1938}
Carruthers,~J.~C. The electrophoresis of certain hydrocarbons and their simple
  derivatives as a function of p. \emph{Trans. Faraday Soc.} \textbf{1938},
  \emph{34}, 300--307\relax
\mciteBstWouldAddEndPuncttrue
\mciteSetBstMidEndSepPunct{\mcitedefaultmidpunct}
{\mcitedefaultendpunct}{\mcitedefaultseppunct}\relax
\EndOfBibitem
\bibitem[Takahashi(2005)]{takahashi2005}
Takahashi,~M. ζ Potential of Microbubbles in Aqueous Solutions:  Electrical
  Properties of the Gas−Water Interface. \emph{The Journal of Physical
  Chemistry B} \textbf{2005}, \emph{109}, 21858--21864, PMID: 16853839\relax
\mciteBstWouldAddEndPuncttrue
\mciteSetBstMidEndSepPunct{\mcitedefaultmidpunct}
{\mcitedefaultendpunct}{\mcitedefaultseppunct}\relax
\EndOfBibitem
\bibitem[Saykally(2013)]{Saykally2013}
Saykally,~R.~J. Two sides of the acid-base story. \emph{Nature Chemistry}
  \textbf{2013}, \emph{5}, 82 EP --\relax
\mciteBstWouldAddEndPuncttrue
\mciteSetBstMidEndSepPunct{\mcitedefaultmidpunct}
{\mcitedefaultendpunct}{\mcitedefaultseppunct}\relax
\EndOfBibitem
\bibitem[Gray-Weale and Beattie(2009)Gray-Weale, and Beattie]{wealebeattie2009}
Gray-Weale,~A.; Beattie,~J.~K. An explanation for the charge on water’s
  surface. \emph{Phys. Chem. Chem. Phys.} \textbf{2009}, \emph{11},
  10994--11005\relax
\mciteBstWouldAddEndPuncttrue
\mciteSetBstMidEndSepPunct{\mcitedefaultmidpunct}
{\mcitedefaultendpunct}{\mcitedefaultseppunct}\relax
\EndOfBibitem
\bibitem[Beattie \latin{et~al.}(2009)Beattie, Djerdjev, and
  Warr]{beattiefaraday2009}
Beattie,~J.~K.; Djerdjev,~A.~M.; Warr,~G.~G. The surface of neat water is
  basic. \emph{Faraday Discuss.} \textbf{2009}, \emph{141}, 31--39\relax
\mciteBstWouldAddEndPuncttrue
\mciteSetBstMidEndSepPunct{\mcitedefaultmidpunct}
{\mcitedefaultendpunct}{\mcitedefaultseppunct}\relax
\EndOfBibitem
\bibitem[Winter \latin{et~al.}(2009)Winter, Faubel, Vácha, and
  Jungwirth]{jungwirth2009}
Winter,~B.; Faubel,~M.; Vácha,~R.; Jungwirth,~P. Reply to comments on
  Frontiers Article ‘Behavior of hydroxide at the water/vapor interface’.
  \emph{Chemical Physics Letters} \textbf{2009}, \emph{481}, 19 -- 21\relax
\mciteBstWouldAddEndPuncttrue
\mciteSetBstMidEndSepPunct{\mcitedefaultmidpunct}
{\mcitedefaultendpunct}{\mcitedefaultseppunct}\relax
\EndOfBibitem
\bibitem[Mishra \latin{et~al.}(2012)Mishra, Enami, Nielsen, Stewart, Hoffmann,
  Goddard, and Colussi]{Mishra2012}
Mishra,~H.; Enami,~S.; Nielsen,~R.~J.; Stewart,~L.~A.; Hoffmann,~M.~R.;
  Goddard,~W.~A.; Colussi,~A.~J. Br{\o}nsted basicity of the
  air{\textendash}water interface. \emph{Proceedings of the National Academy of
  Sciences} \textbf{2012}, \emph{109}, 18679--18683\relax
\mciteBstWouldAddEndPuncttrue
\mciteSetBstMidEndSepPunct{\mcitedefaultmidpunct}
{\mcitedefaultendpunct}{\mcitedefaultseppunct}\relax
\EndOfBibitem
\bibitem[Petersen and Saykally(2005)Petersen, and
  Saykally]{PetersenSaykally2005}
Petersen,~P.~B.; Saykally,~R.~J. Evidence for an Enhanced Hydronium
  Concentration at the Liquid Water Surface. \emph{The Journal of Physical
  Chemistry B} \textbf{2005}, \emph{109}, 7976--7980, PMID: 16851932\relax
\mciteBstWouldAddEndPuncttrue
\mciteSetBstMidEndSepPunct{\mcitedefaultmidpunct}
{\mcitedefaultendpunct}{\mcitedefaultseppunct}\relax
\EndOfBibitem
\bibitem[Petersen and Saykally(2008)Petersen, and Saykally]{petersen2008}
Petersen,~P.~B.; Saykally,~R.~J. Is the liquid water surface basic or acidic?
  Macroscopic vs. molecular-scale investigations. \emph{Chemical Physics
  Letters} \textbf{2008}, \emph{458}, 255 -- 261\relax
\mciteBstWouldAddEndPuncttrue
\mciteSetBstMidEndSepPunct{\mcitedefaultmidpunct}
{\mcitedefaultendpunct}{\mcitedefaultseppunct}\relax
\EndOfBibitem
\bibitem[Tian \latin{et~al.}(2008)Tian, Ji, Waychunas, and Shen]{Shen2008}
Tian,~C.; Ji,~N.; Waychunas,~G.~A.; Shen,~Y.~R. Interfacial Structures of
  Acidic and Basic Aqueous Solutions. \emph{Journal of the American Chemical
  Society} \textbf{2008}, \emph{130}, 13033--13039, PMID: 18774819\relax
\mciteBstWouldAddEndPuncttrue
\mciteSetBstMidEndSepPunct{\mcitedefaultmidpunct}
{\mcitedefaultendpunct}{\mcitedefaultseppunct}\relax
\EndOfBibitem
\bibitem[Tarbuck \latin{et~al.}(2006)Tarbuck, Ota, and Richmond]{tarbuck2006}
Tarbuck,~T.~L.; Ota,~S.~T.; Richmond,~G.~L. Spectroscopic Studies of Solvated
  Hydrogen and Hydroxide Ions at Aqueous Surfaces. \emph{Journal of the
  American Chemical Society} \textbf{2006}, \emph{128}, 14519--14527, PMID:
  17090035\relax
\mciteBstWouldAddEndPuncttrue
\mciteSetBstMidEndSepPunct{\mcitedefaultmidpunct}
{\mcitedefaultendpunct}{\mcitedefaultseppunct}\relax
\EndOfBibitem
\bibitem[Gopalakrishnan \latin{et~al.}(2006)Gopalakrishnan, Liu, Allen, Kuo,
  and Shultz]{Shultz2006}
Gopalakrishnan,~S.; Liu,~D.; Allen,~H.~C.; Kuo,~M.; Shultz,~M.~J. Vibrational
  Spectroscopic Studies of Aqueous Interfaces:  Salts, Acids, Bases, and
  Nanodrops. \emph{Chemical Reviews} \textbf{2006}, \emph{106}, 1155--1175,
  PMID: 16608176\relax
\mciteBstWouldAddEndPuncttrue
\mciteSetBstMidEndSepPunct{\mcitedefaultmidpunct}
{\mcitedefaultendpunct}{\mcitedefaultseppunct}\relax
\EndOfBibitem
\bibitem[Yan \latin{et~al.}(2018)Yan, Delgado, Aubry, Gribelin, Stocco,
  Boisson-Da~Cruz, Bernard, and Ganachaud]{Ganachaud2018}
Yan,~X.; Delgado,~M.; Aubry,~J.; Gribelin,~O.; Stocco,~A.; Boisson-Da~Cruz,~F.;
  Bernard,~J.; Ganachaud,~F. Central Role of Bicarbonate Anions in Charging
  Water/Hydrophobic Interfaces. \emph{The Journal of Physical Chemistry
  Letters} \textbf{2018}, \emph{9}, 96--103, PMID: 29239612\relax
\mciteBstWouldAddEndPuncttrue
\mciteSetBstMidEndSepPunct{\mcitedefaultmidpunct}
{\mcitedefaultendpunct}{\mcitedefaultseppunct}\relax
\EndOfBibitem
\bibitem[Okur \latin{et~al.}(2018)Okur, Drexler, Tyrode, Cremer, and
  Roke]{okur2018}
Okur,~H.~I.; Drexler,~C.~I.; Tyrode,~E.; Cremer,~P.~S.; Roke,~S. The
  Jones–Ray Effect Is Not Caused by Surface-Active Impurities. \emph{The
  Journal of Physical Chemistry Letters} \textbf{2018}, \emph{9}, 6739--6743,
  PMID: 30398354\relax
\mciteBstWouldAddEndPuncttrue
\mciteSetBstMidEndSepPunct{\mcitedefaultmidpunct}
{\mcitedefaultendpunct}{\mcitedefaultseppunct}\relax
\EndOfBibitem
\bibitem[Buch \latin{et~al.}(2007)Buch, Milet, V{\'a}cha, Jungwirth, and
  Devlin]{Buch2007}
Buch,~V.; Milet,~A.; V{\'a}cha,~R.; Jungwirth,~P.; Devlin,~J.~P. Water surface
  is acidic. \emph{Proceedings of the National Academy of Sciences}
  \textbf{2007}, \emph{104}, 7342--7347\relax
\mciteBstWouldAddEndPuncttrue
\mciteSetBstMidEndSepPunct{\mcitedefaultmidpunct}
{\mcitedefaultendpunct}{\mcitedefaultseppunct}\relax
\EndOfBibitem
\bibitem[Baer \latin{et~al.}(2014)Baer, Kuo, Tobias, and Mundy]{Baer2014}
Baer,~M.~D.; Kuo,~I.-F.~W.; Tobias,~D.~J.; Mundy,~C.~J. Toward a Unified
  Picture of the Water Self-Ions at the Air–Water Interface: A Density
  Functional Theory Perspective. \emph{The Journal of Physical Chemistry B}
  \textbf{2014}, \emph{118}, 8364--8372, PMID: 24762096\relax
\mciteBstWouldAddEndPuncttrue
\mciteSetBstMidEndSepPunct{\mcitedefaultmidpunct}
{\mcitedefaultendpunct}{\mcitedefaultseppunct}\relax
\EndOfBibitem
\bibitem[Mundy \latin{et~al.}(2009)Mundy, Kuo, Tuckerman, Lee, and
  Tobias]{mundy2009}
Mundy,~C.~J.; Kuo,~I.-F.~W.; Tuckerman,~M.~E.; Lee,~H.-S.; Tobias,~D.~J.
  Hydroxide anion at the air–water interface. \emph{Chemical Physics Letters}
  \textbf{2009}, \emph{481}, 2 -- 8\relax
\mciteBstWouldAddEndPuncttrue
\mciteSetBstMidEndSepPunct{\mcitedefaultmidpunct}
{\mcitedefaultendpunct}{\mcitedefaultseppunct}\relax
\EndOfBibitem
\bibitem[Petersen \latin{et~al.}(2004)Petersen, Iyengar, Day, and Voth]{voth1}
Petersen,~M.~K.; Iyengar,~S.~S.; Day,~T. J.~F.; Voth,~G.~A. The Hydrated Proton
  at the Water Liquid/Vapor Interface. \emph{The Journal of Physical Chemistry
  B} \textbf{2004}, \emph{108}, 14804--14806\relax
\mciteBstWouldAddEndPuncttrue
\mciteSetBstMidEndSepPunct{\mcitedefaultmidpunct}
{\mcitedefaultendpunct}{\mcitedefaultseppunct}\relax
\EndOfBibitem
\bibitem[Tse \latin{et~al.}(2015)Tse, Chen, Lindberg, Kumar, and Voth]{voth2}
Tse,~Y.-L.~S.; Chen,~C.; Lindberg,~G.~E.; Kumar,~R.; Voth,~G.~A. Propensity of
  Hydrated Excess Protons and Hydroxide Anions for the Air–Water Interface.
  \emph{Journal of the American Chemical Society} \textbf{2015}, \emph{137},
  12610--12616, PMID: 26366480\relax
\mciteBstWouldAddEndPuncttrue
\mciteSetBstMidEndSepPunct{\mcitedefaultmidpunct}
{\mcitedefaultendpunct}{\mcitedefaultseppunct}\relax
\EndOfBibitem
\bibitem[Kumar \latin{et~al.}(2013)Kumar, Knight, and Voth]{voth3}
Kumar,~R.; Knight,~C.; Voth,~G.~A. Exploring the behaviour of the hydrated
  excess proton at hydrophobic interfaces. \emph{Faraday Discuss.}
  \textbf{2013}, \emph{167}, 263--278\relax
\mciteBstWouldAddEndPuncttrue
\mciteSetBstMidEndSepPunct{\mcitedefaultmidpunct}
{\mcitedefaultendpunct}{\mcitedefaultseppunct}\relax
\EndOfBibitem
\bibitem[Mamatkulov \latin{et~al.}(2017)Mamatkulov, Allolio, Netz, and
  Bonthuis]{Netz2017}
Mamatkulov,~S.~I.; Allolio,~C.; Netz,~R.~R.; Bonthuis,~D.~J.
  Orientation-Induced Adsorption of Hydrated Protons at the Air–Water
  Interface. \emph{Angewandte Chemie International Edition} \textbf{2017},
  \emph{56}, 15846--15851\relax
\mciteBstWouldAddEndPuncttrue
\mciteSetBstMidEndSepPunct{\mcitedefaultmidpunct}
{\mcitedefaultendpunct}{\mcitedefaultseppunct}\relax
\EndOfBibitem
\bibitem[Gasparotto \latin{et~al.}(2016)Gasparotto, Hassanali, and
  Ceriotti]{gasparotto2016}
Gasparotto,~P.; Hassanali,~A.~A.; Ceriotti,~M. Probing Defects and Correlations
  in the Hydrogen-Bond Network of ab Initio Water. \emph{Journal of Chemical
  Theory and Computation} \textbf{2016}, \emph{12}, 1953--1964, PMID:
  26881726\relax
\mciteBstWouldAddEndPuncttrue
\mciteSetBstMidEndSepPunct{\mcitedefaultmidpunct}
{\mcitedefaultendpunct}{\mcitedefaultseppunct}\relax
\EndOfBibitem
\bibitem[Vácha \latin{et~al.}(2012)Vácha, Marsalek, Willard, Bonthuis, Netz,
  and Jungwirth]{jungwirth2012}
Vácha,~R.; Marsalek,~O.; Willard,~A.~P.; Bonthuis,~D.~J.; Netz,~R.~R.;
  Jungwirth,~P. Charge Transfer between Water Molecules As the Possible Origin
  of the Observed Charging at the Surface of Pure Water. \emph{The Journal of
  Physical Chemistry Letters} \textbf{2012}, \emph{3}, 107--111\relax
\mciteBstWouldAddEndPuncttrue
\mciteSetBstMidEndSepPunct{\mcitedefaultmidpunct}
{\mcitedefaultendpunct}{\mcitedefaultseppunct}\relax
\EndOfBibitem
\bibitem[Wick \latin{et~al.}(2012)Wick, Lee, and Rick]{wickrick2012}
Wick,~C.~D.; Lee,~A.~J.; Rick,~S.~W. How intermolecular charge transfer
  influences the air-water interface. \emph{The Journal of Chemical Physics}
  \textbf{2012}, \emph{137}, 154701\relax
\mciteBstWouldAddEndPuncttrue
\mciteSetBstMidEndSepPunct{\mcitedefaultmidpunct}
{\mcitedefaultendpunct}{\mcitedefaultseppunct}\relax
\EndOfBibitem
\bibitem[Samson \latin{et~al.}(2014)Samson, Scheu, Smolentsev, Rick, and
  Roke]{samson2014}
Samson,~J.-S.; Scheu,~R.; Smolentsev,~N.; Rick,~S.~W.; Roke,~S. Sum frequency
  spectroscopy of the hydrophobic nanodroplet/water interface: Absence of
  hydroxyl ion and dangling OH bond signatures. \emph{Chemical Physics Letters}
  \textbf{2014}, \emph{615}, 124 -- 131\relax
\mciteBstWouldAddEndPuncttrue
\mciteSetBstMidEndSepPunct{\mcitedefaultmidpunct}
{\mcitedefaultendpunct}{\mcitedefaultseppunct}\relax
\EndOfBibitem
\bibitem[Björneholm \latin{et~al.}(2016)Björneholm, Hansen, Hodgson, Liu,
  Limmer, Michaelides, Pedevilla, Rossmeisl, Shen, Tocci, Tyrode, Walz, Werner,
  and Bluhm]{hendrik2016}
Björneholm,~O.; Hansen,~M.~H.; Hodgson,~A.; Liu,~L.-M.; Limmer,~D.~T.;
  Michaelides,~A.; Pedevilla,~P.; Rossmeisl,~J.; Shen,~H.; Tocci,~G.;
  Tyrode,~E.; Walz,~M.-M.; Werner,~J.; Bluhm,~H. Water at Interfaces.
  \emph{Chemical Reviews} \textbf{2016}, \emph{116}, 7698--7726, PMID:
  27232062\relax
\mciteBstWouldAddEndPuncttrue
\mciteSetBstMidEndSepPunct{\mcitedefaultmidpunct}
{\mcitedefaultendpunct}{\mcitedefaultseppunct}\relax
\EndOfBibitem
\bibitem[Griffith and Vaida(2012)Griffith, and Vaida]{Griffith2012}
Griffith,~E.~C.; Vaida,~V. In situ observation of peptide bond formation at the
  water{\textendash}air interface. \emph{Proceedings of the National Academy of
  Sciences} \textbf{2012}, \emph{109}, 15697--15701\relax
\mciteBstWouldAddEndPuncttrue
\mciteSetBstMidEndSepPunct{\mcitedefaultmidpunct}
{\mcitedefaultendpunct}{\mcitedefaultseppunct}\relax
\EndOfBibitem
\bibitem[Mompe{\'a}n \latin{et~al.}(2019)Mompe{\'a}n, Mar{\'i}n-Yaseli,
  Espigares, Gonz{\'a}lez-Toril, Zorzano, and Ruiz-Bermejo]{Mompean2019}
Mompe{\'a}n,~C.; Mar{\'i}n-Yaseli,~M.~R.; Espigares,~P.;
  Gonz{\'a}lez-Toril,~E.; Zorzano,~M.-P.; Ruiz-Bermejo,~M. Prebiotic chemistry
  in neutral/reduced-alkaline gas-liquid interfaces. \emph{Scientific Reports}
  \textbf{2019}, \emph{9}, 1916\relax
\mciteBstWouldAddEndPuncttrue
\mciteSetBstMidEndSepPunct{\mcitedefaultmidpunct}
{\mcitedefaultendpunct}{\mcitedefaultseppunct}\relax
\EndOfBibitem
\bibitem[Lin \latin{et~al.}(2013)Lin, Cheng, Lin, Lee, and Wang]{Lin2013}
Lin,~Z.-H.; Cheng,~G.; Lin,~L.; Lee,~S.; Wang,~Z.~L. Water–Solid Surface
  Contact Electrification and its Use for Harvesting Liquid-Wave Energy.
  \emph{Angewandte Chemie} \textbf{2013}, \emph{125}, 12777--12781\relax
\mciteBstWouldAddEndPuncttrue
\mciteSetBstMidEndSepPunct{\mcitedefaultmidpunct}
{\mcitedefaultendpunct}{\mcitedefaultseppunct}\relax
\EndOfBibitem
\bibitem[Baytekin \latin{et~al.}(2011)Baytekin, Patashinski, Branicki,
  Baytekin, Soh, and Grzybowski]{Baytekin2011}
Baytekin,~H.~T.; Patashinski,~A.~Z.; Branicki,~M.; Baytekin,~B.; Soh,~S.;
  Grzybowski,~B.~A. The Mosaic of Surface Charge in Contact Electrification.
  \emph{Science} \textbf{2011}, \emph{333}, 308--312\relax
\mciteBstWouldAddEndPuncttrue
\mciteSetBstMidEndSepPunct{\mcitedefaultmidpunct}
{\mcitedefaultendpunct}{\mcitedefaultseppunct}\relax
\EndOfBibitem
\bibitem[Abraham \latin{et~al.}(2015)Abraham, Murtola, Schulz, Páll, Smith,
  Hess, and Lindahl]{GROMA}
Abraham,~M.~J.; Murtola,~T.; Schulz,~R.; Páll,~S.; Smith,~J.~C.; Hess,~B.;
  Lindahl,~E. GROMACS: High performance molecular simulations through
  multi-level parallelism from laptops to supercomputers. \emph{SoftwareX}
  \textbf{2015}, \emph{1-2}, 19 -- 25\relax
\mciteBstWouldAddEndPuncttrue
\mciteSetBstMidEndSepPunct{\mcitedefaultmidpunct}
{\mcitedefaultendpunct}{\mcitedefaultseppunct}\relax
\EndOfBibitem
\bibitem[Abascal and Vega(2005)Abascal, and Vega]{TIP-pap}
Abascal,~J. L.~F.; Vega,~C. A general purpose model for the condensed phases of
  water: TIP4P/2005. \emph{The Journal of Chemical Physics} \textbf{2005},
  \emph{123}, 234505\relax
\mciteBstWouldAddEndPuncttrue
\mciteSetBstMidEndSepPunct{\mcitedefaultmidpunct}
{\mcitedefaultendpunct}{\mcitedefaultseppunct}\relax
\EndOfBibitem
\bibitem[Siu \latin{et~al.}(2012)Siu, Pluhackova, and Böckmann]{OIL-FF}
Siu,~S. W.~I.; Pluhackova,~K.; Böckmann,~R.~A. Optimization of the OPLS-AA
  Force Field for Long Hydrocarbons. \emph{Journal of Chemical Theory and
  Computation} \textbf{2012}, \emph{8}, 1459--1470, PMID: 26596756\relax
\mciteBstWouldAddEndPuncttrue
\mciteSetBstMidEndSepPunct{\mcitedefaultmidpunct}
{\mcitedefaultendpunct}{\mcitedefaultseppunct}\relax
\EndOfBibitem
\bibitem[Parrinello and Rahman(1981)Parrinello, and Rahman]{Barost}
Parrinello,~M.; Rahman,~A. Polymorphic transitions in single crystals: A new
  molecular dynamics method. \emph{Journal of Applied Physics} \textbf{1981},
  \emph{52}, 7182--7190\relax
\mciteBstWouldAddEndPuncttrue
\mciteSetBstMidEndSepPunct{\mcitedefaultmidpunct}
{\mcitedefaultendpunct}{\mcitedefaultseppunct}\relax
\EndOfBibitem
\bibitem[Alejandre and Chapela(2010)Alejandre, and Chapela]{SurfTens}
Alejandre,~J.; Chapela,~G.~A. The surface tension of TIP4P/2005 water model
  using the Ewald sums for the dispersion interactions. \emph{The Journal of
  Chemical Physics} \textbf{2010}, \emph{132}, 014701\relax
\mciteBstWouldAddEndPuncttrue
\mciteSetBstMidEndSepPunct{\mcitedefaultmidpunct}
{\mcitedefaultendpunct}{\mcitedefaultseppunct}\relax
\EndOfBibitem
\bibitem[Skylaris \latin{et~al.}(2005)Skylaris, Haynes, Mostofi, and
  Payne]{ONET}
Skylaris,~C.-K.; Haynes,~P.~D.; Mostofi,~A.~A.; Payne,~M.~C. Introducing
  ONETEP: Linear-scaling density functional simulations on parallel computers.
  \emph{The Journal of Chemical Physics} \textbf{2005}, \emph{122},
  084119\relax
\mciteBstWouldAddEndPuncttrue
\mciteSetBstMidEndSepPunct{\mcitedefaultmidpunct}
{\mcitedefaultendpunct}{\mcitedefaultseppunct}\relax
\EndOfBibitem
\bibitem[Bowler and Miyazaki(2012)Bowler, and Miyazaki]{LS-Rev}
Bowler,~D.~R.; Miyazaki,~T. {\textbackslash}mathcal$\lbrace$O$\rbrace$(N)
  methods in electronic structure calculations. \emph{Reports on Progress in
  Physics} \textbf{2012}, \emph{75}, 036503\relax
\mciteBstWouldAddEndPuncttrue
\mciteSetBstMidEndSepPunct{\mcitedefaultmidpunct}
{\mcitedefaultendpunct}{\mcitedefaultseppunct}\relax
\EndOfBibitem
\bibitem[Prodan and Kohn(2005)Prodan, and Kohn]{Nearsight}
Prodan,~E.; Kohn,~W. Nearsightedness of electronic matter. \emph{Proceedings of
  the National Academy of Sciences of the United States of America}
  \textbf{2005}, \emph{102}, 11635--11638\relax
\mciteBstWouldAddEndPuncttrue
\mciteSetBstMidEndSepPunct{\mcitedefaultmidpunct}
{\mcitedefaultendpunct}{\mcitedefaultseppunct}\relax
\EndOfBibitem
\bibitem[Cloizeaux(1964)]{DM1}
Cloizeaux,~J.~D. Energy Bands and Projection Operators in a Crystal: Analytic
  and Asymptotic Properties. \emph{Phys. Rev.} \textbf{1964}, \emph{135},
  A685--A697\relax
\mciteBstWouldAddEndPuncttrue
\mciteSetBstMidEndSepPunct{\mcitedefaultmidpunct}
{\mcitedefaultendpunct}{\mcitedefaultseppunct}\relax
\EndOfBibitem
\bibitem[Ismail-Beigi and Arias(1999)Ismail-Beigi, and Arias]{DM2}
Ismail-Beigi,~S.; Arias,~T.~A. Locality of the Density Matrix in Metals,
  Semiconductors, and Insulators. \emph{Phys. Rev. Lett.} \textbf{1999},
  \emph{82}, 2127--2130\relax
\mciteBstWouldAddEndPuncttrue
\mciteSetBstMidEndSepPunct{\mcitedefaultmidpunct}
{\mcitedefaultendpunct}{\mcitedefaultseppunct}\relax
\EndOfBibitem
\bibitem[McWeeny(1960)]{DM3}
McWeeny,~R. Some Recent Advances in Density Matrix Theory. \emph{Rev. Mod.
  Phys.} \textbf{1960}, \emph{32}, 335--369\relax
\mciteBstWouldAddEndPuncttrue
\mciteSetBstMidEndSepPunct{\mcitedefaultmidpunct}
{\mcitedefaultendpunct}{\mcitedefaultseppunct}\relax
\EndOfBibitem
\bibitem[Hernandez and Gillan(1995)Hernandez, and Gillan]{DM4}
Hernandez,~E.; Gillan,~M.~J. Self-consistent first-principles technique with
  linear scaling. \emph{Phys. Rev. B} \textbf{1995}, \emph{51},
  10157--10160\relax
\mciteBstWouldAddEndPuncttrue
\mciteSetBstMidEndSepPunct{\mcitedefaultmidpunct}
{\mcitedefaultendpunct}{\mcitedefaultseppunct}\relax
\EndOfBibitem
\bibitem[Skylaris \latin{et~al.}(2002)Skylaris, Mostofi, Haynes, Di\'eguez, and
  Payne]{NGWF}
Skylaris,~C.-K.; Mostofi,~A.~A.; Haynes,~P.~D.; Di\'eguez,~O.; Payne,~M.~C.
  Nonorthogonal generalized Wannier function pseudopotential plane-wave method.
  \emph{Phys. Rev. B} \textbf{2002}, \emph{66}, 035119\relax
\mciteBstWouldAddEndPuncttrue
\mciteSetBstMidEndSepPunct{\mcitedefaultmidpunct}
{\mcitedefaultendpunct}{\mcitedefaultseppunct}\relax
\EndOfBibitem
\bibitem[He and Vanderbilt(2001)He, and Vanderbilt]{DM-BG}
He,~L.; Vanderbilt,~D. Exponential Decay Properties of Wannier Functions and
  Related Quantities. \emph{Phys. Rev. Lett.} \textbf{2001}, \emph{86},
  5341--5344\relax
\mciteBstWouldAddEndPuncttrue
\mciteSetBstMidEndSepPunct{\mcitedefaultmidpunct}
{\mcitedefaultendpunct}{\mcitedefaultseppunct}\relax
\EndOfBibitem
\bibitem[VandeVondele \latin{et~al.}(2005)VandeVondele, Krack, Mohamed,
  Parrinello, Chassaing, and Hutter]{cp2kqui}
VandeVondele,~J.; Krack,~M.; Mohamed,~F.; Parrinello,~M.; Chassaing,~T.;
  Hutter,~J. Quickstep: Fast and accurate density functional calculations using
  a mixed Gaussian and plane waves approach. \emph{Computer Physics
  Communications} \textbf{2005}, \emph{167}, 103 -- 128\relax
\mciteBstWouldAddEndPuncttrue
\mciteSetBstMidEndSepPunct{\mcitedefaultmidpunct}
{\mcitedefaultendpunct}{\mcitedefaultseppunct}\relax
\EndOfBibitem
\bibitem[Becke(1988)]{B88pap}
Becke,~A.~D. Density-functional exchange-energy approximation with correct
  asymptotic behavior. \emph{Phys. Rev. A} \textbf{1988}, \emph{38},
  3098--3100\relax
\mciteBstWouldAddEndPuncttrue
\mciteSetBstMidEndSepPunct{\mcitedefaultmidpunct}
{\mcitedefaultendpunct}{\mcitedefaultseppunct}\relax
\EndOfBibitem
\bibitem[Lee \latin{et~al.}(1988)Lee, Yang, and Parr]{LYPpap}
Lee,~C.; Yang,~W.; Parr,~R.~G. Development of the Colle-Salvetti
  correlation-energy formula into a functional of the electron density.
  \emph{Phys. Rev. B} \textbf{1988}, \emph{37}, 785--789\relax
\mciteBstWouldAddEndPuncttrue
\mciteSetBstMidEndSepPunct{\mcitedefaultmidpunct}
{\mcitedefaultendpunct}{\mcitedefaultseppunct}\relax
\EndOfBibitem
\bibitem[Grimme(2006)]{GD2}
Grimme,~S. Semiempirical GGA-type density functional constructed with a
  long-range dispersion correction. \emph{Journal of Computational Chemistry}
  \textbf{2006}, \emph{27}, 1787--1799\relax
\mciteBstWouldAddEndPuncttrue
\mciteSetBstMidEndSepPunct{\mcitedefaultmidpunct}
{\mcitedefaultendpunct}{\mcitedefaultseppunct}\relax
\EndOfBibitem
\bibitem[Kleinman and Bylander(1982)Kleinman, and Bylander]{KLpseudo}
Kleinman,~L.; Bylander,~D.~M. Efficacious Form for Model Pseudopotentials.
  \emph{Phys. Rev. Lett.} \textbf{1982}, \emph{48}, 1425--1428\relax
\mciteBstWouldAddEndPuncttrue
\mciteSetBstMidEndSepPunct{\mcitedefaultmidpunct}
{\mcitedefaultendpunct}{\mcitedefaultseppunct}\relax
\EndOfBibitem
\bibitem[opi()]{opiumsite}
\url{http://opium.sourceforge.net/sci.html}\relax
\mciteBstWouldAddEndPuncttrue
\mciteSetBstMidEndSepPunct{\mcitedefaultmidpunct}
{\mcitedefaultendpunct}{\mcitedefaultseppunct}\relax
\EndOfBibitem
\bibitem[Manz and Sholl(2012)Manz, and Sholl]{DDECbas}
Manz,~T.~A.; Sholl,~D.~S. Improved Atoms-in-Molecule Charge Partitioning
  Functional for Simultaneously Reproducing the Electrostatic Potential and
  Chemical States in Periodic and Nonperiodic Materials. \emph{Journal of
  Chemical Theory and Computation} \textbf{2012}, \emph{8}, 2844--2867, PMID:
  26592125\relax
\mciteBstWouldAddEndPuncttrue
\mciteSetBstMidEndSepPunct{\mcitedefaultmidpunct}
{\mcitedefaultendpunct}{\mcitedefaultseppunct}\relax
\EndOfBibitem
\bibitem[Lee \latin{et~al.}(2014)Lee, Limas, Cole, Payne, Skylaris, and
  Manz]{ONEDDEC}
Lee,~L.~P.; Limas,~N.~G.; Cole,~D.~J.; Payne,~M.~C.; Skylaris,~C.-K.;
  Manz,~T.~A. Expanding the Scope of Density Derived Electrostatic and Chemical
  Charge Partitioning to Thousands of Atoms. \emph{Journal of Chemical Theory
  and Computation} \textbf{2014}, \emph{10}, 5377--5390, PMID: 26583221\relax
\mciteBstWouldAddEndPuncttrue
\mciteSetBstMidEndSepPunct{\mcitedefaultmidpunct}
{\mcitedefaultendpunct}{\mcitedefaultseppunct}\relax
\EndOfBibitem
\bibitem[Becke(1993)]{B3LYPpap}
Becke,~A.~D. Density‐functional thermochemistry. III. The role of exact
  exchange. \emph{The Journal of Chemical Physics} \textbf{1993}, \emph{98},
  5648--5652\relax
\mciteBstWouldAddEndPuncttrue
\mciteSetBstMidEndSepPunct{\mcitedefaultmidpunct}
{\mcitedefaultendpunct}{\mcitedefaultseppunct}\relax
\EndOfBibitem
\bibitem[M\o{}ller and Plesset(1934)M\o{}ller, and Plesset]{MP2pap}
M\o{}ller,~C.; Plesset,~M.~S. Note on an Approximation Treatment for
  Many-Electron Systems. \emph{Phys. Rev.} \textbf{1934}, \emph{46},
  618--622\relax
\mciteBstWouldAddEndPuncttrue
\mciteSetBstMidEndSepPunct{\mcitedefaultmidpunct}
{\mcitedefaultendpunct}{\mcitedefaultseppunct}\relax
\EndOfBibitem
\bibitem[Babin \latin{et~al.}(2013)Babin, Leforestier, and Paesani]{MBPOLpap}
Babin,~V.; Leforestier,~C.; Paesani,~F. Development of a “First Principles”
  Water Potential with Flexible Monomers: Dimer Potential Energy Surface, VRT
  Spectrum, and Second Virial Coefficient. \emph{Journal of Chemical Theory and
  Computation} \textbf{2013}, \emph{9}, 5395--5403, PMID: 26592277\relax
\mciteBstWouldAddEndPuncttrue
\mciteSetBstMidEndSepPunct{\mcitedefaultmidpunct}
{\mcitedefaultendpunct}{\mcitedefaultseppunct}\relax
\EndOfBibitem
\bibitem[Willard and Chandler(2010)Willard, and Chandler]{Istant-surf}
Willard,~A.~P.; Chandler,~D. Instantaneous Liquid Interfaces. \emph{The Journal
  of Physical Chemistry B} \textbf{2010}, \emph{114}, 1954--1958\relax
\mciteBstWouldAddEndPuncttrue
\mciteSetBstMidEndSepPunct{\mcitedefaultmidpunct}
{\mcitedefaultendpunct}{\mcitedefaultseppunct}\relax
\EndOfBibitem
\bibitem[Agmon \latin{et~al.}(2016)Agmon, Bakker, Campen, Henchman, Pohl, Roke,
  Thämer, and Hassanali]{agmon2016}
Agmon,~N.; Bakker,~H.~J.; Campen,~R.~K.; Henchman,~R.~H.; Pohl,~P.; Roke,~S.;
  Thämer,~M.; Hassanali,~A. Protons and Hydroxide Ions in Aqueous Systems.
  \emph{Chemical Reviews} \textbf{2016}, \emph{116}, 7642--7672, PMID:
  27314430\relax
\mciteBstWouldAddEndPuncttrue
\mciteSetBstMidEndSepPunct{\mcitedefaultmidpunct}
{\mcitedefaultendpunct}{\mcitedefaultseppunct}\relax
\EndOfBibitem
\bibitem[Giberti and Hassanali(2017)Giberti, and
  Hassanali]{gibertihassanali2017}
Giberti,~F.; Hassanali,~A.~A. The excess proton at the air-water interface: The
  role of instantaneous liquid interfaces. \emph{The Journal of Chemical
  Physics} \textbf{2017}, \emph{146}, 244703\relax
\mciteBstWouldAddEndPuncttrue
\mciteSetBstMidEndSepPunct{\mcitedefaultmidpunct}
{\mcitedefaultendpunct}{\mcitedefaultseppunct}\relax
\EndOfBibitem
\bibitem[Hsieh \latin{et~al.}(2011)Hsieh, Campen, Vila~Verde, Bolhuis,
  Nienhuys, and Bonn]{mischabonn2011}
Hsieh,~C.-S.; Campen,~R.~K.; Vila~Verde,~A.~C.; Bolhuis,~P.; Nienhuys,~H.-K.;
  Bonn,~M. Ultrafast Reorientation of Dangling OH Groups at the Air-Water
  Interface Using Femtosecond Vibrational Spectroscopy. \emph{Phys. Rev. Lett.}
  \textbf{2011}, \emph{107}, 116102\relax
\mciteBstWouldAddEndPuncttrue
\mciteSetBstMidEndSepPunct{\mcitedefaultmidpunct}
{\mcitedefaultendpunct}{\mcitedefaultseppunct}\relax
\EndOfBibitem
\bibitem[K{\"u}hne and Khaliullin(2013)K{\"u}hne, and Khaliullin]{Kuhne2013}
K{\"u}hne,~T.~D.; Khaliullin,~R.~Z. Electronic signature of the instantaneous
  asymmetry in the first coordination shell of liquid water. \emph{Nature
  Communications} \textbf{2013}, \emph{4}, 1450 EP --, Article\relax
\mciteBstWouldAddEndPuncttrue
\mciteSetBstMidEndSepPunct{\mcitedefaultmidpunct}
{\mcitedefaultendpunct}{\mcitedefaultseppunct}\relax
\EndOfBibitem
\bibitem[Khaliullin \latin{et~al.}(2009)Khaliullin, Bell, and
  Head-Gordon]{bellkhaliullin2009}
Khaliullin,~R.; Bell,~A.; Head-Gordon,~M. Electron Donation in the
  Water–Water Hydrogen Bond. \emph{Chemistry – A European Journal}
  \textbf{2009}, \emph{15}, 851--855\relax
\mciteBstWouldAddEndPuncttrue
\mciteSetBstMidEndSepPunct{\mcitedefaultmidpunct}
{\mcitedefaultendpunct}{\mcitedefaultseppunct}\relax
\EndOfBibitem
\bibitem[Phipps \latin{et~al.}(2017)Phipps, Fox, Tautermann, and
  Skylaris]{EDA-ON}
Phipps,~M. J.~S.; Fox,~T.; Tautermann,~C.~S.; Skylaris,~C.-K. Intuitive Density
  Functional Theory-Based Energy Decomposition Analysis for Protein–Ligand
  Interactions. \emph{Journal of Chemical Theory and Computation}
  \textbf{2017}, \emph{13}, 1837--1850\relax
\mciteBstWouldAddEndPuncttrue
\mciteSetBstMidEndSepPunct{\mcitedefaultmidpunct}
{\mcitedefaultendpunct}{\mcitedefaultseppunct}\relax
\EndOfBibitem
\bibitem[Winter \latin{et~al.}(2009)Winter, Faubel, Vácha, and
  Jungwirth]{winterjungwirth2009}
Winter,~B.; Faubel,~M.; Vácha,~R.; Jungwirth,~P. Behavior of hydroxide at the
  water/vapor interface. \emph{Chemical Physics Letters} \textbf{2009},
  \emph{474}, 241 -- 247\relax
\mciteBstWouldAddEndPuncttrue
\mciteSetBstMidEndSepPunct{\mcitedefaultmidpunct}
{\mcitedefaultendpunct}{\mcitedefaultseppunct}\relax
\EndOfBibitem
\end{mcitethebibliography}

\end{document}